# Birthday attack to discrete logarithm

Li An-Ping

Beijing 100080, P.R. China
apli0001@sina.com

**Abstract:** The discrete logarithm in a finite group of large order has been widely applied in public key cryptosystem. In this paper, we will present a probabilistic algorithm for discrete logarithm.

## 1. Introduction

The generalized Discrete Logarithm problem (GDLP) has been taken as an intractable problem widely applied in public key cryptosystems since it firstly was proposed by W. Diffie and M.E. Hellman in their key exchange system [1]. The problem is the one to find the solution for the exponential equation $a^x = b$ in a group $G$. Since then, there have been several algorithms for GDLP presented, in which main three ones are that, the baby-step giant-step algorithm, Silver- Pohlig- Hellman's algorithm and the index calculus algorithm. Suppose that the order of $G$ $|G|=N$, then the required operations of the first algorithm will take about $O(\sqrt{N})$ group multiplications and $O(\sqrt{N} \cdot \log N)$ comparisons. The second one is similar to the method of solving the congruence in number theory by resolve the equation into the subgroups of group $G$ with the order of prime powers, so it is more efficient if the prime factors of group $G$ all are smaller. In the case the base groups $G$ are from the multiplication groups of finite fields, the index calculus is the most powerful one. For the detail about GDLP, refer to see [2] and A.M. Odlyzko's survey papers [3], [4].

In this paper, we will present a probabilistic algorithm for discrete logarithm, which is a development of the birthday attack.

As usual, for two integers $x$ and $r$, we will represent $[x]_r = x \cdot (x-1) \cdots (x-r+1)$, and $[x]^r = x \cdot (x+1) \cdots (x+r-1)$. Let $A$ be a set of $t$ elements, in which elements repetitions are allowed, suppose that there are just $t-d$ different elements in $A$, then we call $A$ has $d$ collisions.

## 2. The description and analysis

Suppose that $G$ is the base group of discrete logarithm, the order of $G$, $|G|=N$, and $g \in G$ is a generator of $G$. Let $S = \{y_i = g^{x_i} \mid 1 \leq i \leq m\}$ be a set of $m$ public keys. Denoted by $n = \lceil N^{1/2} \rceil + 1$, for each $y_i \in S, 1 \leq i \leq m$, arbitrary takes $n$ distinct integers $r_j^{(i)}, 1 \leq j \leq n$, and makes a set

$$S_i = \{ y_i^{r_j^{(i)}} \mid 1 \leq j \leq n \}, 1 \leq i \leq m.$$

Moreover, arbitrary takes $n$ distinct integers $r_j^{(0)}, 1 \leq j \leq n$, and makes a set

$$S_0 = \{ g^{r_j^{(i)}} \mid 1 \leq j \leq n \}.$$

**Proposition 1** Let $\Omega = \bigcup_{0 \leq i \leq m} S_i$, suppose that $k$ is a non-negative integer, $k \leq m$, denoted by $p_k(\Omega)$ (or simply, $p_k$) the probability that there are at least $k$ collisions in the set $\Omega$, it has

$$p_k \geq 1 - e^{-(m+1)^2 + \varepsilon} \cdot (m+1)^{2k} / k!. \qquad (2.1)$$

where $\varepsilon \prec 2(T-k)^3 / 3N^2$.

Proof. Let $T = |\Omega| = (m+1)n$, for each integer $i, 0 \leq i < m$, denoted by $\lambda_i$ the number of all possible $T$-subsets of $N$ distinct elements with $i$ collisions, then it is easy to know that

$$\lambda_i = C_N^{T-i} \cdot [T-i]^i / i! = C_N^{T-i} \cdot C_{T-1}^i,$$

Hence,

$$\begin{aligned}
p_k &\geq 1 - \sum_{0 \leq i < k} \lambda_i / ([N]^T / T!) \geq 1 - \sum_{0 \leq i < k} [N]_{T-i} \cdot [T]_i \cdot C_{T-1}^i / [N]^T \\
&\geq 1 - \prod_{1 \leq i < T-k} (1 - \frac{2i}{N+i}) \cdot \sum_{0 \leq i < k} (m+1)^{2i} / i! \\
&\geq 1 - \exp(\sum_{1 \leq i < T-k} \frac{-2i}{N+i})(m+1)^{2k} / k! \\
&\geq 1 - \exp(\sum_{1 \leq i < T-k} \frac{-2i}{N} + \frac{2i^2}{N^2})(m+1)^{2k} / k! \\
&\geq 1 - e^{-(m+1)^2 + 2(T-k)^3 / 3N^2} \cdot (m+1)^{2k} / k!.
\end{aligned}$$

$\square$

From the proposition above, we know that the probability $p_m(\Omega)$ will be greater than $0.99$ as $m \geq 2$.

Clearly, a collision in the set $\Omega$ is just a linear equation of the privacy keys $\{x_i\}_1^m$, say in detail, if $i, j \neq 0$

$$y_i^{r_s^{(i)}} = y_j^{r_t^{(j)}} \Leftrightarrow r_s^{(i)} \cdot x_i = r_t^{(j)} \cdot x_j \mod N,$$

and

$$y_i^{r_s^{(i)}} = y_j^{r_t^{(0)}} \Leftrightarrow r_s^{(i)} \cdot x_i = r_t^{(0)} \mod N.$$

So, in the case that these $m$ linear equations are linear independent, then all the privacy keys $\{x_i\}_1^m$ will be discovered by solving the obtained linear system.

On the other hand, we know the probability that randomly take $m$ $m$-vectors which are linear independent is very great, which is about $1 - (1/N)$.

It is clear that the main computations required in this new analysis are the comparisons to find $m$ collisions in the set $\Omega$ and the group multiplications in making the sets $S_i$. By the simple way that classifying the elements of $\Omega$ in bits one by one, the required comparisons to order the elements in $\Omega$ are about $O(T \log T)$, which is about equal to the required comparisons in the baby-step giant-step algorithm. If properly selecting the constants $r_j^{(i)}$, the amount of group multiplications required in the new algorithm also will be about equal to the one of the baby-step giant-step algorithm. However, in the case that $m$ is greater, we can reduce the size $n$ of the sets $S_i$ into $n = \left\lceil \left(\frac{2N}{m+1}\right)^{1/2} \right\rceil + 1$, then we have the following estimation

$$\begin{aligned} p_m(\Omega) &\geq 1 - e^{-2(m+1)+\varepsilon} \cdot (2m+2)^m / m! \\ &\geq 1 - \frac{e^{-1+\varepsilon}}{\sqrt{2\pi m}} \left(\frac{2}{e}\right)^m. \end{aligned} \quad (2.2)$$

So, in this way, the required group multiplications in the presented algorithm will be less than the ones in the baby-step giant-step algorithm.

Finally, similar to Silver- Pohlig- Hellman's algorithm, we can resolve the original GDLP into the subgroups of the cyclic group generated by element $g$, and then apply this analysis above to the subgroups.